\documentclass[conference]{IEEEtran}
\IEEEoverridecommandlockouts

\usepackage[T1]{fontenc}
\usepackage{amsmath,amssymb}
\usepackage{graphicx}
\usepackage{booktabs}
\usepackage{multirow}
\usepackage{array}
\usepackage{xcolor}
\usepackage{listings}
\usepackage{algorithm}
\usepackage{algpseudocode}
\usepackage{microtype}
\usepackage{cite}
\usepackage{braket}
\usepackage[hidelinks]{hyperref}

\lstdefinelanguage{QDSL}{
  keywordstyle = \ttfamily
  morekeywords={PREPARE,SUPERPOSE,ENTANGLE,MEASURE,USE,OBSERVABLE,ASSERT,OPTIMIZE,BLOCK,REPEAT,INSPECT_IR},
  sensitive=true,
  morecomment=[l]{\#},
  morestring=[b]"
}
\title{Intent-Level Quantum Programming with Assertion-Guided Execution and Inspectable Intermediate Representations}

\author{
\IEEEauthorblockN{
Ilesh Vora\IEEEauthorrefmark{1},
Srikanth Thudumu\IEEEauthorrefmark{1},
John Carlson\IEEEauthorrefmark{2},
Harshil Kamdar\IEEEauthorrefmark{3}, and
Rajesh Vasa\IEEEauthorrefmark{4}
}
\IEEEauthorblockA{
\IEEEauthorrefmark{1}\textit{Institute of Applied Artificial Intelligence and Robotics (IAAIR), USA}\\
\IEEEauthorrefmark{2}\textit{Stanford University, USA}\\
\IEEEauthorrefmark{3}\textit{Rutgers University, USA}\\
\IEEEauthorrefmark{4}\textit{Applied Artificial Intelligence Initiative
(\ensuremath{A^{2}I^{2}}), Deakin University, Australia}\\[2pt]
\texttt{ilesh.vora@iaair.ai; srikanth@iaair.ai; jwc10@stanford.edu;}\\
\texttt{harshil.kamdar@rutgers.edu; rajesh.vasa@deakin.edu.au}
}
}

\begin{document}
\maketitle

\begin{abstract}
Quantum programs are difficult to validate: circuits are typically expressed as imperative gate sequences with limited inspectability, execution modalities must be selected manually, and outputs are inherently probabilistic. These challenges are compounded when programs must be portable across backend frameworks with differing internal conventions. We present a quantum domain-specific language (QDSL) that addresses these problems through three mechanisms: (i) intent-execution separation, where algorithmic constructs such as preparation, superposition, entanglement, and measurement are represented as inspectable objects compiled into backend-specific circuits only after pre-execution Intermediate Representation (IR) validation; (ii) first-class IR introspection, exposing circuit width, wire mapping, operation order, and measurement intent for developer inspection prior to execution; and (iii) assertion-guided modality inference, where the engine examines declared verification properties to automatically select sampling, statevector, or dual execution without user intervention. Results are logged in structured form to support reproducible regression testing. We evaluate our prototype on benchmark circuits spanning entanglement, oracle-based, structured-transform, and variational examples. In fault-injection experiments on Bell and 3-qubit GHZ circuits, a total-variation-distance detector identifies the injected faults across all evaluated shot budgets (True Positive Rate 1.0), with observed false positives (False Positive Rate 2.7--3.8\%) only at 128 shots. Differential testing demonstrates numerical agreement across PennyLane and Qiskit compilation targets for the evaluated constructs after endianness canonicalization. IR generation costs are sub-millisecond. 

\end{abstract}


\section{Introduction}
Programming quantum algorithms remains challenging in the noisy intermediate-scale quantum (NISQ) era because developers must reason simultaneously about algorithmic structure, circuit-level constraints, and backend-specific execution semantics \cite{preskill2018nisq}. Mainstream frameworks such as PennyLane \cite{bergholm2018pennylane}, Qiskit \cite{qiskit2019zenodo}, and Cirq \cite{cirq2025zenodo} provide powerful abstractions and, in fact, expose rich inspection facilities. 

Our observation is not that inspection is unavailable, but that it is typically imperative-first: intent is expressed as gate sequences, and inspection is offered at the circuit level after construction rather than at the intent level at which the developer was reasoning. For some developers and workflows this gap adds cognitive overhead and complicates debugging, particularly when outputs are probabilistic and measurement is destructive.

A second, closely related challenge is testing. Prior work has proposed statistical assertions for quantum programs \cite{huang2019statassert} and projection-based runtime assertions \cite{li2020projection}, as well as metamorphic and differential testing approaches \cite{wang2021qdiff,paltenghi2023morphq}. These efforts highlight the need for developer-facing verification tools, especially given the prevalence of quantum-specific bugs in platforms and toolchains \cite{paltenghi2022bugs}. However, existing workflows often leave developers to manually choose execution modalities (shot-based vs.\ statevector), wire test harnesses, and interpret intermediate circuit artifacts. We consolidate language design and verification into a single framework: an intent-centric QDSL whose construction layer is intent-named and whose verification layer is declarative, paired with Intermediate Representation (IR) introspection and an assertion-guided verification engine. Designed around software-engineering mental models, it separates (i) intent expression, (ii) compilation and introspection, and (iii) execution and evaluation. This separation reduces ambiguity between circuit construction and execution phases, enables pre-execution IR validation, and supports a coherent debugging loop.

The contributions of this paper are:
\begin{enumerate}
  \item \textbf{Intent-execution separation via intent-named construction.} We introduce a QDSL that treats program intent (e.g., \texttt{PREPARE}, \texttt{SUPERPOSE}, \texttt{ENTANGLE}, \texttt{MEASURE}, \texttt{USE}) as first-class data that can be inspected and validated prior to compilation, rather than as immediately executed host-language calls.
  \item \textbf{First-class IR introspection for quantum debugging.} We define a user-facing IR that exposes circuit width, wire usage, operation order, parameters, and measurement/observable intent, enabling inspection. Comparable circuit-level information is already available post-construction in existing frameworks (e.g., PennyLane's \texttt{specs} and \texttt{draw}). Our distinction: inspection occurs at the intent level, before backend binding, with each operation tied via an \texttt{origin} field to the reusable \texttt{BLOCK} that produced it.
  \item \textbf{Assertion-guided modality inference.} We implement an assertion engine that combines declared assertions with output requirements recorded in the IR to determine whether sampling, statevector, or mixed execution is required, reporting \mbox{PASS/FAIL} outcomes using statistically meaningful checks.
  \item \textbf{Phase-aware API semantics.} We enforce a framework-agnostic Type A/B/C taxonomy separating circuit builders (Type A), analyzers (Type B), and runners (Type C) to prevent common build-vs-run confusion and enable phase-aware validation.
\end{enumerate}

We integrate compilation, inspection, fault injection, repeated trials, and structured logging into a single workflow suitable for regression testing and continuous integration (CI).

\section{Related Work}

Higher-level quantum languages such as Quipper \cite{green2013quipper} and Silq \cite{bichsel2020silq} explore new abstractions and type systems; we instead target an intent-centric QDSL embedding that treats verification as a first-class feature and emphasizes developer workflows. For validation, prior work has proposed statistical \cite{huang2019statassert} and projection-based runtime assertions \cite{li2020projection}, and differential \cite{wang2021qdiff} and metamorphic testing \cite{paltenghi2023morphq,abreu2022metamorphic}, motivated by the prevalence of quantum-specific bugs \cite{paltenghi2022bugs}. Our approach unifies language-level intent representation with IR introspection and assertion-guided execution, and explicitly addresses execution-modality selection and reproducible artifact generation.

\section{Approach}
\subsection{System overview}
\begin{figure*}[t]
  \centering
  \IfFileExists{change_backend_fleet.png}{%
    \includegraphics[width=0.85\textwidth]{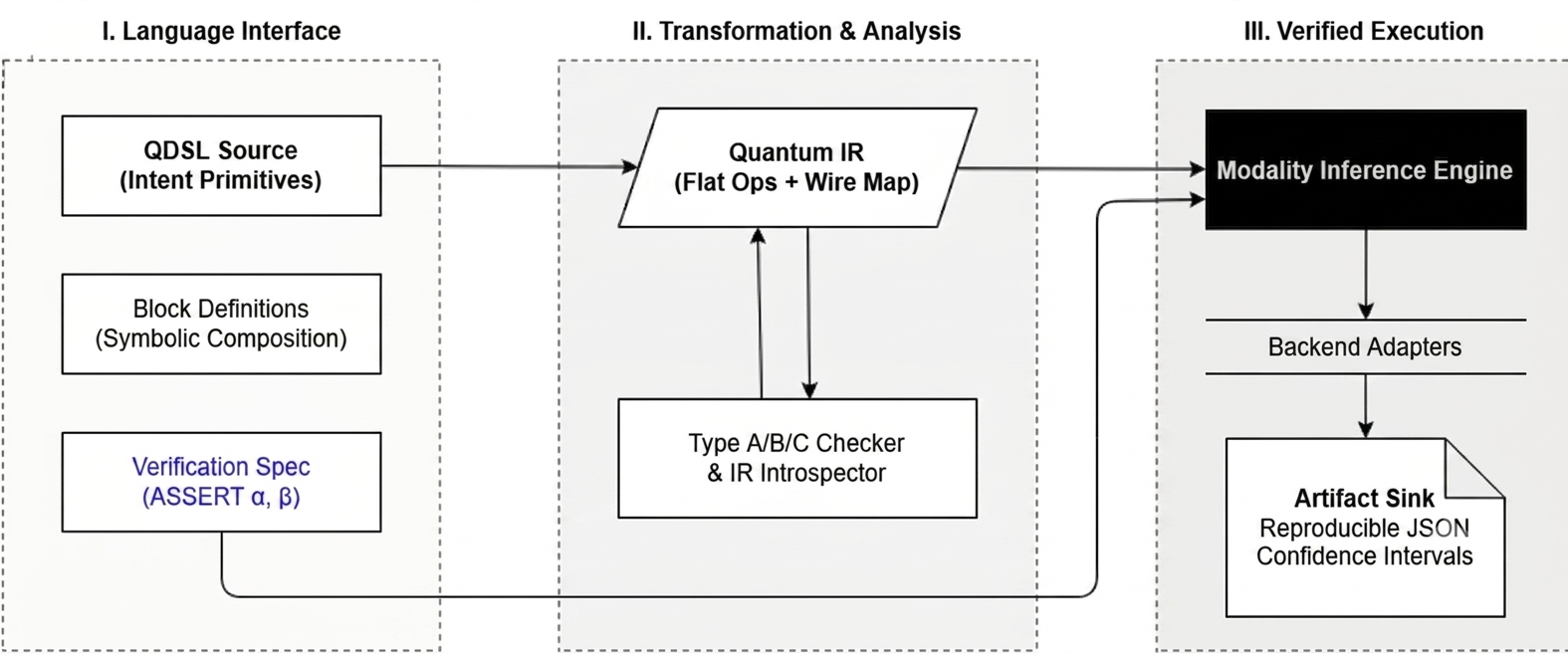}%
  }{%
    \fbox{\parbox{1.3\textwidth}{\centering Missing file: qdsl-ieee.png}}%
  }
  \caption{QDSL system overview and compilation/verification workflow.}
  \label{fig:qdsl_system_overview}
\end{figure*}

Figure~\ref{fig:qdsl_system_overview} summarizes the workflow. A program is authored in QDSL, compiled into an inspectable IR, validated before execution (including Type A/B/C constraints), optionally mutated for fault injection, and then bound to one or more backends. Assertions guide execution modality selection. Results are aggregated across repetitions and emitted as structured logs.

\begin{lstlisting}[caption={Illustrative QDSL surface syntax for a Bell-state program},captionpos=b, label={lst:qdsl-bell}]
with PREPARE(2) as p:
     SUPERPOSE(0)
     ENTANGLE(0,1)
     MEASURE q --> c
ASSERT prob(c == "00") == 0.5
ASSERT prob(c == "11") == 0.5
\end{lstlisting}

\subsection{Intent-named construction and reusable blocks}
The QDSL encodes intent as named constructs aligned with a developer's mental model of the algorithm. Listing~\ref{lst:qdsl-bell} shows a minimal Bell-state program expressed at an algorithmic level without gate-by-gate plumbing. For probability assertions, the equality operator denotes statistical compatibility with the specified probability under the configured confidence interval, rather than exact equality of the observed sample frequency. By default, QDSL evaluates such assertions using a 95\% Wilson score interval.

Instead of executing operations immediately, QDSL constructs an intent graph that is lowered into IR. Reusable \texttt{BLOCK}s are first-class QDSL objects (registered, parameterizable, composable), enabling algorithm-level composition while preserving inspectability.

We note that \texttt{ENTANGLE} currently targets a fixed, common set of entangling patterns (pairwise CNOT-style entanglement and a built-in family of named Bell-state blocks covering all four Bell states, $\ket{\Phi^\pm}$ and $\ket{\Psi^\pm}$), rather than an open family of arbitrary entangled states. This is a deliberate scoping choice for the present work: the construct covers the entangling structure required by our benchmarks while keeping intent unambiguous. Richer entanglement specifications are a natural extension and are left to future work.

\subsection{IR introspection as a debugging and pedagogical tool}

\begin{table}[t]
\centering
\caption{User-facing IR fields exposed for inspection.}
\label{tab:irfields}
\begin{tabular}{@{}p{0.25\columnwidth}p{0.6\columnwidth}@{}}
\toprule
\textbf{IR field} & \textbf{Meaning} \\
\midrule
Circuit width & Total number of logical qubits required \\
Wire map & Mapping from source-level qubit identifiers to canonical logical
wire indices\\
Ordered ops & Linearized operation sequence \\
Parameters & Symbolic and numeric parameters for gates/observables \\
Origin & Reusable \texttt{BLOCK} each operation came from \\
Measurement intent & Which wires are measured and in what basis \\
Observable intent & Observables for expectation values / fidelities \\
\bottomrule
\end{tabular}
\end{table}

Table~\ref{tab:irfields} lists the IR fields exposed by the inspection interface. The \texttt{INSPECT\_IR} implementation was aligned to Table~\ref{tab:irfields}.

To make this concrete, we inspect a two-qubit program that prepares a Bell state, applies an \texttt{RZ} rotation, and measures the Pauli-Z expectation value on the first qubit. Crucially, the IR is available before the program is bound to any backend, so the developer can confirm circuit structure independently of PennyLane or Qiskit conventions. \texttt{INSPECT\_IR} emits a machine-readable dictionary by default (Listing~\ref{lst:ir-dict}, \textbf{Appendix~\ref{app:ir-dict}}) to support automation and regression, while \texttt{INSPECT\_IR(p,"text")} renders a human-readable view for debugging (Listing~\ref{lst:ir-text}).

\subsection{Phase-aware semantics via a Type A/B/C taxonomy}
A frequent source of confusion in decorator-style quantum APIs is the boundary between circuit construction and execution. We classify QDSL-exposed functions into: (i) \textbf{Type A} builders (construct circuits/IR), (ii) \textbf{Type B} analyzers (inspect IR without mutation or execution), and (iii) \textbf{Type C} runners (execute on a backend). To illustrate the value of this taxonomy, consider a common mistake: a developer might attempt to apply a gate directly in the main execution flow rather than inside a defined block or preparation context. Under QDSL's phase semantics, calling a Type A builder like \texttt{SUPERPOSE} outside of a \texttt{PREPARE} block immediately raises a runtime error (e.g., \texttt{ContextError: SUPERPOSE must be called within a valid PREPARE context}). These boundaries are enforced by phase-aware runtime checks at construction time, before backend binding, eliminating silent state-mutations, reducing semantic misuse, and enabling predictable debugging workflows. 

\begin{lstlisting}[caption={Human-readable IR (\texttt{text} format).}, captionpos=b, label={lst:ir-text}]
Circuit width: 2
Wire map: {'q[0]': 0, 'q[1]': 1}
Ordered ops:
  1. H(wires=[0], params=[]) origin=['Rotate_Bell']
  2. CNOT(wires=[0, 1], params=[]) origin=['Rotate_Bell']
  3. RZ(wires=[0], params=[0.78539816339744]) origin=['Rotate_Bell']
  4. MEASURE(kind=expval, wires=[0], basis=Z)

Measurement intent:
  kind=expval wires=[0] basis=Z

Observable intent:
  {'type': 'Z', 'wires': [0]}
\end{lstlisting}
\subsection{Assertion-guided execution via modality inference}
Quantum program outputs are probabilistic; consequently, validation must reason about uncertainty. The framework supports both sampling-based and statevector-based assertions. For sampling assertions, we compute confidence intervals for binomial proportions (by default Wilson score intervals \cite{wilson1927}, with optional exact intervals \cite{clopper1934}). For state-based assertions, fidelities or amplitudes are evaluated directly; expectation values may be evaluated by sampling or statevector depending on the assertion (Table~\ref{tab:modality_mapping}; Algorithm~\ref{alg:modality_inference}). A key design principle is \emph{modality inference}: the engine examines declared assertions and automatically selects the minimal required execution mode. This prevents modality mismatches and supports mixed-mode verification in a single run.

\begin{table}[h]
\centering
\caption{Mapping of assertion property groups to execution modalities.}
\label{tab:modality_mapping}
\setlength{\tabcolsep}{4pt}
\renewcommand{\arraystretch}{1.05}
\begin{tabular}{@{}p{0.23\columnwidth}p{0.50\columnwidth}p{0.2\columnwidth}@{}}
\toprule
\textbf{Property group} & \textbf{Examples} & \textbf{Modality ($\mathcal{M}$)} \\ \midrule
Algebraic state & Fidelity, amplitudes, trace distance & Statevector \\
Statistical output & Bitstring probabilities, counts & Sampling \\
Observables & Pauli expectations & Samp./Statev. \\
Structural & Width, depth, op counts & Static (IR only) \\
\bottomrule
\end{tabular}
\end{table}

\begin{algorithm}[h!]
\caption{Assertion-guided execution modality inference}
\label{alg:modality_inference}
\small
\begin{algorithmic}[1]
\Require Circuit IR $\mathcal{I}$ and assertion set
         $\mathcal{A}=\{a_1,\ldots,a_n\}$
\Ensure Execution modality
        $\mathcal{M}\in\{\text{Static},\text{Statevector},
        \text{Sampling},\text{Dual}\}$
\State $\mathcal{S}\gets\emptyset$
\For{each assertion $a\in\mathcal{A}$}
    \If{\Call{RequiresStatevector}{$a$}}
        \State $\mathcal{S}\gets
        \mathcal{S}\cup\{\text{Statevector}\}$
    \ElsIf{\Call{RequiresSampling}{$a$}}
        \State $\mathcal{S}\gets
        \mathcal{S}\cup\{\text{Sampling}\}$
    \ElsIf{\Call{IsExpectation}{$a$}}
        \If{$a.\text{modality}\notin
            \{\text{Statevector},\text{Sampling}\}$}
            \State \textbf{raise} InvalidAssertionModality
        \EndIf
        \State $\mathcal{S}\gets
        \mathcal{S}\cup\{a.\text{modality}\}$
    \ElsIf{\Call{IsStructural}{$a$}}
        \State \textbf{continue}
        \Comment{Evaluated directly on the IR}
    \Else
        \State \textbf{raise} UnsupportedAssertionTarget
    \EndIf
\EndFor
\If{$\mathcal{S}=\emptyset$}
    \State \Return $\text{Static}$
\ElsIf{$|\mathcal{S}|=1$}
    \State \Return the only element of $\mathcal{S}$
\Else
    \State \Return $\text{Dual}$
\EndIf

\end{algorithmic}
\end{algorithm}
\textsc{RequiresStatevector} covers fidelity, statevector,
amplitude, and trace-distance assertions, while
\textsc{RequiresSampling} covers probability, count, sampled-output,
and distribution-distance assertions. Expectation assertions specify
their required modality, whereas structural assertions are evaluated
directly on the IR. Declared output requests are resolved from the IR
before being combined with assertion requirements.

\subsection{Reference-based verification and the oracle requirement}
In our setup, the framework does not infer user intent from scratch. Instead, each benchmark test is anchored by an explicit reference specification: (i) a trusted PennyLane circuit, i.e. an independently written implementation whose correctness is established outside the QDSL toolchain (by construction from the textbook algorithm and prior validation); (ii) a golden QDSL program, i.e. a previously verified QDSL implementation whose output is treated as the expected baseline for regression testing; or (iii) a stored IR snapshot capturing the expected intermediate representation. When a user selects a benchmark test (e.g., ``run GHZ''), they implicitly select that reference. The framework compiles the current candidate implementation, runs candidate and reference under the same shot budget, and checks whether their distributions or higher-level properties agree within tolerance. This design instantiates a practical oracle strategy for quantum program testing, consistent with the broader oracle problem in software testing \cite{barr2015oracle}.

\section{Evaluation}

\subsection{Research Questions \& Metrics}
We structure the evaluation around the following research questions:
\begin{enumerate}
    \item \textbf{RQ1: Source-Level Abstraction.}
How does QDSL affect source-code size and the proportion of code devoted to algorithm-specific intent compared with imperative
implementations?
    \item \textbf{RQ2: Fault Detection Under Sampling Noise.}
    How reliably does the framework detect injected defects when outputs are probabilistic?
    \item \textbf{RQ3: Cross-Backend Agreement.}
    To what extent do compiled programs agree across tooling stacks and execution backends?
    \item \textbf{RQ4: IR Generation Overhead.}
    What computational overhead does the QDSL introduce during frontend preprocessing before backend execution?
\end{enumerate}

\subsection{Benchmarks and baselines}

We implemented the benchmark suite in QDSL, PennyLane and Qiskit, covering Bell States, Greenberger-Horne-Zeilinger (GHZ), Superdense Coding, Swap Test, Quantum Approximate Optimization Algorithm MaxCut (QAOA), Deutsch--Jozsa, Grover search, Quantum Fourier Transform (QFT), Quantum Teleportation, and Variational Quantum Eigensolver (VQE) examples.


\subsection{Experimental Setup}

Experiments were run in a pinned environment: Python 3.12, PennyLane v0.35, Qiskit v1.0. Circuits were executed on local simulators (PennyLane's \texttt{default.qubit}, Qiskit's \texttt{qiskit\_aer.AerSimulator} \cite{qiskit_aer_simulator}); no physical hardware was targeted, with exact statevector simulation for algebraic properties and pseudo-random sampling for statistical outputs. 
RQ2 used shot budgets of $\{128,512,2048,8192\}$ with a fixed master seed and a distinct deterministic seed for each trial and
configuration. Wilson intervals were used for probability-bound assertions, whereas the fault-injection experiment used TVD with
the fixed threshold $\tau=0.1$.

We isolate the QDSL frontend preprocessing cost (RQ4)---the time and peak memory to build and canonicalize the IR prior to backend binding---on an Intel Core i7-12700H with 16GB RAM running Windows 11. Each benchmark was measured over 20 independent repetitions, timing IR construction and canonicalization separately via Python's \texttt{time.perf\_counter()}; peak memory was recorded as the maximum traced allocation during IR generation using \texttt{tracemalloc}. All outcomes (parsed IR artifacts, execution parameters, assertion results) are logged automatically as structured JSON for reproducible regression testing and CI.

\subsection{RQ1: Abstraction effectiveness}
``A good abstraction must expose the features of a system that must be directly addressed, while hiding the rest behind an appropriate interface'' \cite{di2024abstraction}. The QDSL aims to increase the share of code devoted to algorithmic intent rather than framework mechanics, letting developers express circuits at the intent level while deferring backend mechanics to compilation. We evaluate this using pragmatic software-engineering proxies. 
We compare functionally equivalent implementations in PennyLane, Qiskit, and the QDSL across the five benchmarks in Table~\ref{tab:loc-comparison}, treating implementations as \textit{functionally equivalent} if they encode the same algorithm at the same problem size and produce the same declared output under the same evaluation setting.


\paragraph{Effort Metrics}
We use three effort proxies:
\begin{enumerate}
    \item \textbf{Lines of Code (LOC) \& Intent Density:} how much explicit source written, and the share of it expressing the algorithm rather than framework mechanics.
    \item \textbf{Boilerplate Categories:} recurring non-algorithmic scaffolding.
    \item \textbf{Composability:} how easily a reusable subroutine can be defined once and reused or swapped without rewriting the enclosing algorithm.
\end{enumerate}

\subsubsection{LOC and Intent Density}

\paragraph{LOC Methodology}

We use LOC as a proxy for implementation burden, not a claim that fewer lines are better. Because LOC is sensitive to formatting, we apply a fixed counting protocol across frameworks: counting non-empty, non-comment source lines \cite{nguyen2007sloc} that contain executable code or a required declaration:

\begin{itemize}
    \item One statement per line wherever practical.
    \item Shared library/framework imports are excluded since they only reflect environment setup rather than benchmark-specific implementation effort. 
    \item Framework-provided utilities are counted as single statements and are not expanded into their internal implementations when computing LOC (e.g., \texttt{qml.GradientDescentOptimizer} in PennyLane).
    \item Presentation-only and inspection-only statements, such as formatted printing, plotting, circuit drawing, or IR visualizations, are excluded unless they are required to produce the benchmark's declared output object.
\end{itemize}

\paragraph{Intent Density}
We calculate this metric as $\text{Intent Density} = N_\text{intent} \div \text{LOC}$, where $N_\text{intent}$ is the number of counted lines that directly express the quantum algorithm. Intent density is a custom proxy, as we are aware of no established metric capturing the intent-vs-scaffolding distinction for quantum programs. It combines line-based effort accounting with the principle that a good abstraction exposes what must be addressed \cite{di2024abstraction}: the fraction of counted lines devoted to algorithmic intent is then an approximate indicator of abstraction quality, meaningful only in comparison across functionally equivalent implementations, and we report it as an indicative proxy rather than a validated metric. 

Each counted line is classified as \textit{intent} (algorithmic steps or algorithm-specific reusable units, e.g.\ \texttt{def oracle\_x0(...)}, \texttt{@BLOCK("qft")}) or \textit{scaffolding} (framework mechanics, e.g. \texttt{def circuit(...)}, \texttt{@qml.qnode}, backend instantiation).

\begin{table}[t]
\centering
\caption{RQ1 Results: LOC and Intent Density (ID) across QDSL, PennyLane (PL), and Qiskit.}
\label{tab:loc-comparison}
\setlength{\tabcolsep}{3pt}
\renewcommand{\arraystretch}{1.1}
\resizebox{\columnwidth}{!}{%
\begin{tabular}{@{}lcccccc@{}}
\toprule
\textbf{Benchmark} & \textbf{PL LOC} & \textbf{PL ID} & \textbf{Qiskit LOC} & \textbf{Qiskit ID} & \textbf{QDSL LOC} & \textbf{QDSL ID} \\
\midrule
Teleportation          & 15 & 0.67 & 16 & 0.50 & 12 & 0.83 \\
Deutsch--Jozsa (2-bit) & 24 & 0.71 & 21 & 0.67 & 27 & 0.85 \\
Grover's Search (3-q)   & 20 & 0.70 & 20 & 0.65 & 20 & 0.85 \\
QFT ($n$-q)            & 15 & 0.67 & 16 & 0.56 & 12 & 0.83 \\
VQE                    & 21 & 0.43 & 24 & 0.29 & 16 & 0.69 \\
\bottomrule
\end{tabular}}
\end{table}

\paragraph{Analysis}

Table~\ref{tab:loc-comparison} reports LOC and intent density across QDSL, PennyLane, and Qiskit. The QDSL achieves the highest intent density on every benchmark. 

The advantage compounds where QDSL also reduces raw LOC---in Teleportation, QFT, and VQE the implementations are shorter and denser; the effect is largest in VQE, where the imperative baselines need substantial scaffolding for optimizer setup, iterative updates, and result handling that QDSL expresses through \texttt{OPTIMIZE}, \texttt{USE}, and \texttt{MEASURE}.

Intent density also exposes benefits that raw LOC hides. In Grover's Search all three implementations have identical LOC, yet QDSL reaches density 0.85 vs.\ 0.70 (PennyLane) and 0.65 (Qiskit)---a larger share devoted to oracle, diffusion, and measurement structure rather than plumbing. Conversely, in 2-bit Deutsch--Jozsa QDSL is \emph{longer} in raw LOC but denser (0.85 vs.\ 0.71 and 0.67): the extra lines are explicit oracle and block structure, underscoring that the benefit extends beyond shorter programs to a greater share of code reflecting algorithmic steps.

\subsubsection{Boilerplate Categories}
\begin{table*}[h]
\centering
\caption{Boilerplate categories across implementations.}
\label{tab:boilerplate}
\setlength{\tabcolsep}{5pt}
\renewcommand{\arraystretch}{1.2}
\begin{tabular}{llll}
\hline
\textbf{Category} & \textbf{PennyLane} & \textbf{Qiskit} & \textbf{QDSL} \\
\hline
Program/circuit container & QNode/circuit function & \texttt{QuantumCircuit} & \texttt{PREPARE} context \\
Backend/device binding & \texttt{qml.device} & \texttt{AerSimulator} / backend & implicit runtime/backend binding \\
Execution wrapper & \texttt{@qml.qnode} & transpile / run workflow & assertion-guided runner \\
Result declaration/extraction & \texttt{qml.probs} / \texttt{state} / \texttt{expval} & get \texttt{counts} / \texttt{statevector} & \texttt{MEASURE} \\
Optimization loop & optimizer + Python loop & optimizer + Python loop & \texttt{OPTIMIZE} \\
Inspection/visualization & \texttt{qml.draw} + external plotting & \texttt{qc.draw} + external plotting & \texttt{INSPECT\_IR}, \texttt{DRAW}, \texttt{GRAPH} \\
\hline
\end{tabular}
\end{table*}

Distinct API-concept counts are comparable across the three frameworks, so we do not treat them as discriminating; the difference is organizational, with QDSL's vocabulary mapping to the developer's \textit{mental model} rather than spanning backend setup, wrappers, construction, and extraction. In relation to this, boilerplate refers to recurring lines necessary for execution, not directly expressing the algorithm; the intent-density analysis classifies these as scaffolding.

As summarized in Table~\ref{tab:boilerplate}, PennyLane and Qiskit baselines exhibit recurring boilerplate across device setup, execution wrappers, result extraction, optimization scaffolding, and visualization. In the QDSL, these responsibilities are handled implicitly by the runtime or exposed through primitives such as \texttt{PREPARE}, \texttt{USE}, \texttt{MEASURE}, and \texttt{OPTIMIZE}, keeping source focused on algorithm structure. Inspection differs notably: where PennyLane and Qiskit generally rely on Matplotlib for diagrams and plots, the QDSL integrates these via \texttt{DRAW}, \texttt{GRAPH}, and \texttt{INSPECT\_IR}, replacing plotting boilerplate with single commands.

\subsubsection{Composability}

Composability is the ability to build programs by combining reusable subroutines without modifying the surrounding structure---relevant in quantum software, where subcircuits (e.g, oracles, diffusion operators, etc) recur across algorithms. The QDSL supports this through the \texttt{BLOCK}+\texttt{USE} mechanism: subroutines are defined once as named blocks and invoked in multiple contexts without duplicating logic.

This appears across the suite: swappable Deutsch--Jozsa oracles, the Grover oracle and diffusion operator defined once and reused across iterations, and a single VQE ansatz block reused across optimization iterations with different parameters.

Imperative frameworks achieve similar reuse through Python functions; QDSL's advantage is the form it takes: \texttt{BLOCK}+\texttt{USE} makes reusable fragments part of the DSL, so subcircuits are registered in the IR, and composable without the caller managing wire indices or device context.

\subsubsection{Implementation Example}

Listing \ref{lst:teleport_qdsl} presents the QDSL implementation of the coherent deferred-measurement form of the Quantum Teleportation algorithm; PennyLane and Qiskit implementations are in \textbf{Appendix~\ref{app:teleport}}. An extension with explicit mid-circuit measurement and classical control is future work.

\begin{lstlisting}[caption={Quantum Teleportation in QDSL}, captionpos=b, label={lst:teleport_qdsl}]
from dsl import *
import numpy as np

@BLOCK("teleport")
def teleport(state):
    STATE_PREP(state, 0)
    BELL_PHI_PLUS(1,2)
    ENTANGLE(0,1)
    SUPERPOSE(0)
    ENTANGLE(1,2)
    gate.CZ([0,2])

state_to_teleport = np.array([1/np.sqrt(2), 1/np.sqrt(2)])

with PREPARE(3) as p:
    USE("teleport", state=state_to_teleport)
    MEASURE("density matrix", 2)

print(p())
\end{lstlisting}

The implementations show the difference in abstraction level. In PennyLane and Qiskit, the teleportation logic is intertwined with execution infrastructure (device configuration, circuit wrappers, explicit result extraction). The QDSL expresses the protocol through higher-level constructs, \texttt{BELL\_PHI\_PLUS}, \texttt{ENTANGLE}, and \texttt{SUPERPOSE}, that map directly to the conceptual steps of the protocol, while \texttt{@BLOCK}+\texttt{USE} keep algorithm definition separate from execution.

\subsection{RQ2: Fault detection under sampling noise}
\label{sec:rq2_results}

We evaluate whether the framework reliably detects injected defects in entangling circuits under finite-shot noise (Table~\ref{tab:rq2_tvd_results}). On top of the QDSL IR, we apply systematic mutations reporting two fault models: \texttt{mutate\_first\_h\_to\_x} for Bell (initial Hadamard $\rightarrow$ $X$) and \texttt{drop\_last\_entangle} for 3-qubit GHZ (dropping the final entangling gate). Each trial is clean or faulty with probability 0.5; we flag a fault when $\mathrm{TVD}(P,Q) = \frac{1}{2}\sum_x |P(x)-Q(x)| \ge \tau$ between candidate and reference distributions, over 200 trials per setting. Across both benchmarks and all shot budgets, the detector achieves TPR = 1.0 (no missed faults). At 128 shots, we observe small false positives (FPR $\approx 2.7\text{-}3.8\%$); for $\ge 512$ shots we observed none, though this is empirical evidence for the tested fault models rather than a general guarantee. 

\begin{table}[t]
\centering
\caption{RQ2 results: TVD-to-reference detector ($\tau=0.1$), 200 trials per setting.}
\label{tab:rq2_tvd_results}
\setlength{\tabcolsep}{5pt}
\renewcommand{\arraystretch}{1.00}
\begin{tabular}{@{}l l r r r@{}}
\toprule
\textbf{Benchmark} & \textbf{Mutation} & \textbf{Shots} & \textbf{TPR} & \textbf{FPR} \\
\midrule
Bell (2q) & H$\rightarrow$X on first qubit & 128  & 1.00 & 0.027 \\
Bell (2q) & H$\rightarrow$X on first qubit & 512  & 1.00 & 0.000 \\
Bell (2q) & H$\rightarrow$X on first qubit & 2048 & 1.00 & 0.000 \\
Bell (2q) & H$\rightarrow$X on first qubit & 8192 & 1.00 & 0.000 \\
\midrule
GHZ (3q)  & drop last entangling gate      & 128  & 1.00 & 0.038 \\
GHZ (3q)  & drop last entangling gate      & 512  & 1.00 & 0.000 \\
GHZ (3q)  & drop last entangling gate      & 2048 & 1.00 & 0.000 \\
GHZ (3q)  & drop last entangling gate      & 8192 & 1.00 & 0.000 \\
\bottomrule
\end{tabular}
\end{table}

\subsection{RQ3: Cross-backend agreement}
We evaluate whether the QDSL produces numerically consistent outputs across software stacks via differential testing between its PennyLane and Qiskit backends. Each program is constructed once in QDSL, compiled independently to both backends, and checked for numerical agreement. Because the frameworks differ in qubit ordering, outputs are canonicalized to PennyLane's native big-endian convention \cite{Atchade-Adelomou:2023gpw} before comparison. We align global phase using the first non-zero reference amplitude before performing element-wise statevector comparison and computing the maximum absolute difference. The suite covers common sources of cross-backend mismatch, including probability distributions and reversed-wire marginals, multi- and parameterized controlled gates, amplitude and basis-state preparation, Pauli and Hadamard expectation values, and reduced density matrices (Table~\ref{tab:rq3_cross_backend}). After canonicalization, all tested circuits agreed numerically across both backends, with maximum absolute differences at or below $2.22 \times 10^{-16}$. Ordering differences that produced statevector indexing mismatches before normalization vanished once normalized. The pipeline therefore produces numerically consistent outputs
across both compilation targets for the evaluated constructs, with the remaining differences attributable to floating-point
precision.

\begin{table}
\centering
\caption{RQ3 results: Cross-backend numerical agreement between compilation targets.}
\label{tab:rq3_cross_backend}
\setlength{\tabcolsep}{4pt}
\renewcommand{\arraystretch}{1.0}
\resizebox{\columnwidth}{!}{
\begin{tabular}{@{}lllc@{}}
\toprule
\textbf{Test} & \textbf{Output Type} & \textbf{Max Abs Diff} & \textbf{Result} \\
\midrule
Bell probabilities & \texttt{probs} & $0.000$ & PASS \\
Prob subset (reversed wires) & \texttt{probs} & $0.000$ & PASS \\
CTRL-Z (two controls) & \texttt{statevector} & $5.55 \times 10^{-17}$ & PASS \\
CRZ rotation & \texttt{statevector} & $0.000$ & PASS \\
State preparation & \texttt{statevector} & $0.000$ & PASS \\
Basis state initialization & \texttt{statevector} & $0.000$ & PASS \\
Expectation value (Pauli X) & \texttt{expval} & $0.000$ & PASS \\
Expectation value (Hadamard) & \texttt{expval} & $2.22 \times 10^{-16}$ & PASS \\
Density matrix (subsystem) & \texttt{density\_matrix} & $0.000$ & PASS \\
\bottomrule
\end{tabular}}

\vspace{1ex}
\raggedright \footnotesize \textit{Note:} Comparison tolerance set to $\text{atol} = 1.0 \times 10^{-9}$ \& $\text{rtol} = 1.0 \times 10^{-7}$.
\end{table}

\subsection{RQ4: IR generation overhead}
\label{sec:rq4_ir_overhead}

\begin{figure}[h!]
    \centering
    \includegraphics[width=0.93\linewidth]{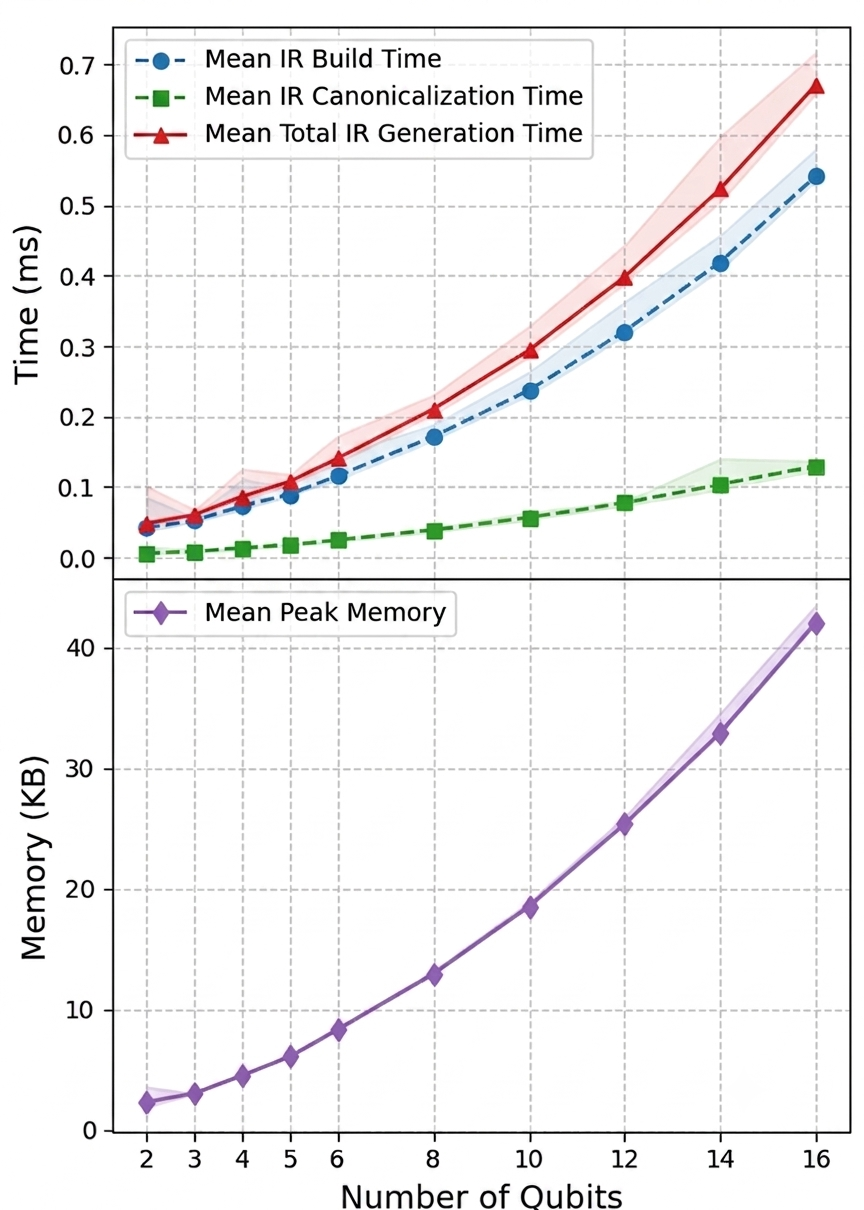}
    \caption{QFT IR preprocessing overhead for $n=2$--$16$. The upper panel reports build, canonicalization, and total preprocessing time; the lower panel reports peak memory. Shaded regions indicate variability across repeated runs.}
    \label{fig:ir_overhead}
\end{figure}

We measure the overhead of building and canonicalizing the QDSL IR (\emph{build+canon}), excluding backend execution, sampling, and assertion evaluation (Table~\ref{tab:rq4_ir_overhead}). 
Build + canon is sub-millisecond for all programs, with peak memory under 11 KB across the benchmark suite (Table~\ref{tab:rq4_ir_overhead}); a QFT scaling study to 
$n=16$ qubits (Figure~\ref{fig:ir_overhead}) shows build time and memory growing with circuit width while remaining small in absolute terms.

\label{app:ir-overhead}

\begin{table}
\centering
\caption{RQ4 Results: IR generation overhead (build + canon). Times are mean ms; memory is mean peak KB.}
\label{tab:rq4_ir_overhead}
\setlength{\tabcolsep}{4pt}
\renewcommand{\arraystretch}{1}
\begin{tabular}{@{}l c c c c c@{}}
\toprule
\textbf{Benchmark} & \textbf{wires} & \textbf{ops} & \textbf{Build} & \textbf{Canon} & \textbf{Peak KB} \\
\midrule
Bell               & 2 &  2 & 0.050 & 0.009 & 1.77 \\
DJ (oracle $x_0$)  & 3 &  7 & 0.108 & 0.021 & 2.83 \\
QFT (4q)           & 4 & 15 & 0.073 & 0.013 & 4.48 \\
Grover's Search             & 3 & 39 & 0.305 & 0.075 & 10.57 \\
VQE (expval)       & 2 &  2 & 0.412 & 0.008 & 4.23 \\
\bottomrule
\end{tabular}\vspace{-5mm}
\end{table}

\section{Discussion}

Two findings stand out: reference-based, distribution-level verification reliably detects the tested faults at $\ge 512$ shots (Table~\ref{tab:rq2_tvd_results}), and IR construction and canonicalization introduce little overhead for the evaluated benchmarks (Table~\ref{tab:rq4_ir_overhead}).
Limitations of this work: our current RQ2 results focus on two fault types and two circuits. Although these are representative of structural defects in entangling programs, broader coverage (e.g., measurement-basis faults, parameter-binding faults, and variational objective faults) is needed. Moreover, reference-based verification depends on a curated oracle (trusted program or property suite) \cite{barr2015oracle}. Finally, metamorphic relations are harder to apply in quantum settings due to probabilistic outputs, destructive measurement, and no-cloning constraints \cite{wootters1982noclone,paltenghi2023morphq,abreu2022metamorphic}; we therefore emphasize statistical, assertion-based validation and differential comparison. A further limitation concerns benchmark scale. Our evaluation uses small, well-understood benchmark circuits because they have unambiguous reference semantics and make the abstraction and verification mechanisms easy to inspect. We therefore treat these as evidence at this scale, not a claim that they generalize unchanged to large, production-scale programs; establishing such generalization is future work.

\section{Conclusion and Future Work}
We presented an intent-centric QDSL with first-class IR introspection and an assertion-guided execution model that selects simulation modalities automatically and produces reproducible logs. The results suggest that integrating intent, inspection, and verification at the language level is a practical path toward more reliable and developer-friendly quantum software. Future work will extend the evaluation in three directions: (i) a developer study measuring time-to-diagnosis with and without IR introspection, (ii) broader fault-injection coverage including measurement-basis and parameter-binding faults, and (iii) expanded cross-backend differential testing as additional backend integrations are completed.

\noindent\textit{Work-in-Progress Statement.}
This work represents an ongoing research effort, and the QDSL
implementation and evaluation remain under active development.
The current results demonstrate the initial feasibility of the
proposed approach rather than a complete or production-ready
system. We expect future versions of this preprint to refine the
language semantics, expand backend support, strengthen the
verification mechanisms, and include broader experimental
evaluation.

\section{Acknowledgments}
We thank Jason Fisher and Ginny Hayes at the Institute of Applied Artificial Intelligence and Robotics (IAAIR) for their support of this work.

\bibliographystyle{IEEEtran}
\bibliography{references}
\appendices


\section{Machine Readable IR Inspection}
\label{app:ir-dict}
\begin{lstlisting}[caption={Machine-readable IR (default \texttt{dict} format).}, captionpos=b, label={lst:ir-dict}]
{ 'Circuit width': 2,
  'Wire map': {'q[0]': 0, 'q[1]': 1},
  'Ordered ops': [ { 'index': 1,
                     'type': 'op',
                     'name': 'H',
                     'wires': [0],
                     'params': [],
                     'origin': ['Rotate_Bell']},
                   { 'index': 2,
                     'type': 'op',
                     'name': 'CNOT',
                     'wires': [0, 1],
                     'params': [],
                     'origin': ['Rotate_Bell']},
                   { 'index': 3,
                     'type': 'op',
                     'name': 'RZ',
                     'wires': [0],
                     'params': [0.78539816339744], 
                     'origin': ['Rotate_Bell']},
                   { 'index': 4,
                     'type': 'measure',
                     'kind': 'expval',
                     'wires': [0],
                     'basis': 'Z',
                     'observable': 'Z',
                     'has_hamiltonian': False,
                     'origin': []}],
  'Parameters': [{'index': 3, 'name': 'RZ', 'params': [0.78539816339744]}],
  'Measurement intent': [{'kind': 'expval', 'wires': [0], 'basis': 'Z'}],
  'Observable intent': [{'type': 'Z', 'wires': [0]}]}
\end{lstlisting}






\vfill\null

\section{PennyLane and Qiskit Implementation of the Quantum Teleportation Algorithm}
\label{app:teleport}






\begin{lstlisting}[caption={Quantum Teleportation in PennyLane}, captionpos=b, label={lst:teleport_pl}]
import pennylane as qml
import numpy as np

def teleport(state):
    qml.StatePrep(state, wires=[0])
    qml.Hadamard(wires=1)
    qml.CNOT(wires=[1, 2])
    qml.CNOT(wires=[0, 1])
    qml.Hadamard(wires=0)
    qml.CNOT(wires=[1, 2])
    qml.CZ(wires=[0, 2])

state_to_teleport = np.array([1/np.sqrt(2), 1/np.sqrt(2)])

dev = qml.device("default.qubit", wires=3)
@qml.qnode(dev)
def circuit():
    teleport(state_to_teleport)
    return qml.density_matrix(2)

rho = circuit()
print(rho)
\end{lstlisting}

\begin{lstlisting}[caption={Quantum Teleportation in Qiskit}, captionpos=b, label={lst:teleport_qiskit}]
import numpy as np
from qiskit import QuantumCircuit
from qiskit.quantum_info import DensityMatrix, partial_trace

def teleport(state):
    qc = QuantumCircuit(3)
    qc.initialize(state, 0)
    qc.h(1)
    qc.cx(1, 2)
    qc.cx(0, 1)
    qc.h(0)
    qc.cx(1, 2)
    qc.cz(0, 2)
    return qc

state_to_teleport = np.array([1/np.sqrt(2), 1/np.sqrt(2)])

def circuit():
    qc = teleport(state_to_teleport)
    rho = partial_trace(DensityMatrix(qc), [0, 1])
    return rho.data

rho = circuit()
print(rho)
\end{lstlisting}


\end{document}